\documentclass[]{JHEP3}
\usepackage{graphicx}
\usepackage{amsmath}
\numberwithin{equation}{section}
\preprint{Cavendish--HEP--11/01}
\title{Eikonal regime of gravity-induced scattering at higher energy proton colliders}
\author{W.J. Stirling, E. Vryonidou, J.D. Wells\\Cavendish Laboratory, J.J. Thomson Avenue, Cambridge CB3 0HE, UK}
\abstract
{We compute transplanckian parton scattering in flat extra-dimensional theories at the LHC and at the recently discussed high-energy upgrade (HE LHC). We report new leading-order calculations of the QCD background. We apply appropriate cuts to satisfy the necessary conditions for the eikonal approximation to be valid while at the same time maximising the signal to background ratio at LHC energies. We study the computability of the eikonal signal and consider the effect of a possible 33 TeV high-energy upgrade to the LHC, which serves to extend the calculable region and to enhance the signal to background ratio.}
\begin{document}

\section{Introduction}
Models with extra dimensions have been proposed to address the hierarchy problem. 
The large extra dimensions model introduced in \cite{ArkaniHamed:1998rs,Antoniadis:1998ig} suggests that the Standard Model (SM) particles live in the usual 3+1-dimensional space, while gravity propagates in a higher D-dimensional space. The weakness of gravity with respect to other forces is imposed by the large size of the compactified extra dimensions. In this model Newton's constant is expressed  as
\begin{equation}
G_N^{-1}=M_{P}^2=8\pi R^{n} M_{D}^{2+n},
\label{Planck}
\end{equation}
where $M_P$ is the Planck mass, $M_D\sim$ TeV is the fundamental mass scale, $n$ is the number of extra dimensions and $R$ the radius of the compactified space, assumed to be a torus in this model. The size of the extra dimensions is determined by their number $n$, and the fundamental mass scale $M_D$, through Eq.~\ref{Planck}. For a fundamental scale of the order of 1~TeV and one extra dimension, its size would be of the order of the solar system size. As this would cause large deviations from the inverse square law of gravitation, one simple extra dimension is already ruled out by experiments. For more extra dimensions, the size gets rapidly smaller, not contradicting submillimetric gravity measurements \cite{Kapner:2006si}.
 
 In the ADD model \cite{ArkaniHamed:1998rs}, the graviton corresponds to the excitations of the D-dimensional metric. These can be expressed as a tower of Kaluza-Klein (KK) modes. The interaction Lagrangian for gravitons and the SM fields is given by
\begin{equation}
\mathcal{L}_{\rm{int}}=-\frac{1}{\overline{M}_{P}}\sum_{\vec{n}}G^{\vec{n}}_{\mu\nu}T^{\mu\nu},
\label{Lag}
\end{equation}
with $\overline{M}_{P}=M_P/\sqrt{8\pi}$ the reduced four-dimensional Planck mass, $T_{\mu\nu}$ the energy momentum tensor of the SM fields and $\vec{n}=(n_1,n_2,...,n_{n})$, a $n$-dimensional vector of integers, labelling the massive gravitons. The interactions of these massive KK states with the Standard Model particles are summarised in a set of Feynman rules presented in \cite{Giudice:1998ck, Han:1998sg}. It is noted that this is an effective linearized theory of interactions valid in the cisplanckian region, that is when the centre-of-mass energy of the parton-parton collision is smaller than the fundamental scale $M_D$. 

The KK resonances have masses equal to $m_{(\vec{n})}=|\vec{n}|/R$. This results in a small mass gap of order $R^{-1}$, e.g. for one extra dimension of size 1~$\mu$m the mass gap is $\mathcal{O}$(1~eV). This renders different masses practically indistinguishable and permits replacing the sum over discrete mass values by an integral over a continuum with a given density of states. This density of states is obtained by considering the number of modes with KK index between $|\vec{n}|$ and $|\vec{n}|+dn$, given by 
\begin{equation}
dN=S_{n-1}|n|^{n-1}dn=S_{n-1}\frac{\overline{M}_{P}^2}{M_D^{2+n}}m^{n-1}dm \,\,\,\mbox{    with    }\,\,\,\,S_{n-1}=\frac{2\pi^{n/2}}{\Gamma(n/2)}.
\end{equation} Thus the density of states is given by
\begin{equation}
\rho(m)=\frac{dN}{dm}=S_{n-1}\frac{\overline{M}_{P}^2}{M_D^{2+n}}m^{n-1}.
\label{density}
\end{equation}

As seen from the Lagrangian in Eq.~\ref{Lag}, graviton interactions are suppressed by inverse powers of $\overline{M}_{P}$. Nevertheless, the summation over the large number of accessible KK modes, that is the integration over the density in Eq.~\ref{density}, cancels the dependence on $\overline{M}_P$. As an example, the inclusive graviton production cross section $\sigma_m$ is expected from the graviton couplings to be proportional to $\overline{M}^{-2}_P$, which combined with the density in Eq.~\ref{density} exactly cancels the dependence on $\overline{M}_P$. This leads to an effective interaction suppressed by inverse powers of the fundamental mass scale $M_D$, thus giving observable effects for $M_D$ near the TeV scale. The result is expected from the higher dimensional field theory perspective.

Interesting phenomenological implications of this extra dimensions model have been studied in the literature over the last decade. 
The processes studied include real graviton emission and virtual graviton exchange. A set of processes, such as graviton plus gauge boson production, was studied in \cite{Giudice:1998ck,Han:1998sg}, where the Feynman rules were first presented. 
The experimental signature for graviton production is missing energy, as decay into SM particles is suppressed by a factor of $1/M_P^2$, which is not compensated for by phase space. Therefore gravitons behave as heavy and stable particles, once produced. This missing-energy signal does not correspond to a fixed invisible-particle mass as the graviton has a continuous mass distribution. This differentiates graviton searches from other Beyond the Standard Model (BSM) physics searches, such as supersymmetry which would also give a missing energy signal.
 The experimental signatures depend strongly on the number of extra dimensions, $n$ and the fundamental mass scale, $M_D$, so in principle one should be able to determine or at least constrain both of these model parameters. 
Virtual graviton exchange effects will also cause deviations from SM predictions for fermion and boson pair production. Single graviton exchange is an observable not fully calculable. The cross sections are divergent at tree level, and thus one has to introduce an ultra-violet (UV) cut-off that is usually taken to be of the order of the fundamental scale \cite{Giudice:1998ck}. The search for virtual graviton effects is complementary to the production search which is reliably calculated when $E\ll M_D$, and will shed light on this cut-off and its dependence on the fundamental scale and the number of extra dimensions.

Both the LEP and Tevatron experiments have set limits on $M_D$.
For a brief summary of the experimental searches and the current constraints on the fundamental scale and the number of extra dimensions see \cite{Landsberg:2008ax} and references therein. A more recent study of the dijet angular distribution by the D0 experiment at the Tevatron was performed in \cite{:2009mh}. The most stringent current constraints on the fundamental scale are set at $1-1.5$~TeV but these also depend on the number of extra dimensions. The LHC is expected to probe extra dimensions effects in all channels mentioned above up to higher scales, due to the larger centre-of-mass energy. Recently, a first study of the bounds set on the cut-off scale from the dijet cross section measured at the LHC was presented in \cite{Franceschini:2011wr}, which are in the few TeV region.
 
The predictions of this effective theory of graviton interactions fail as we approach the quantum gravity scale. As we reach the planckian region where $\sqrt{s}\simeq M_D$, we need theoretical input from quantum gravity, and so it is presently impossible to reliably predict experimental signals. An important feature of the present hadron colliders is that the range of accessible energies potentially allows the simultaneous probe of cisplanckian, planckian and transplanckian regions. Eagerly awaited experimental results from the LHC and future colliders could thus provide crucial information on the quantum gravity region.
 
The transplanckian region, $\sqrt{s}\gg M_D$, can also be studied in a fairly model independent way~\cite{Amati:1992zb,Emparan:2001kf}. In this high energy limit, the two experimental signatures are fermion pair production from elastic parton scattering and black hole production~\cite{Giddings:2001bu,Dimopoulos:2001hw}. Calculation of the characteristic scales of the system for elastic cross section in the transplanckian region \cite{Giudice:2001ce} shows that it can be described by classical physics. Furthermore, short distance gravity effects such as black hole formation are suppressed by considering scattering with large impact parameters. For large impact parameters the curvature is small and so we are in the limit of weak gravitational field. The scattering amplitude can be computed using the eikonal approximation, valid when the scattering angle is small. In this approach, the divergent nature of the theory is overcome by resumming an infinite number of ladder and cross-ladder Feynman diagrams which give the leading contribution to forward scattering \cite{Abarbanel:1969ek}. This has been studied in \cite{Giudice:2001ce}. Black hole production is also briefly studied in \cite{Giudice:2001ce} in this kinematical region, within a dimensional argument approach. High energy gravitational scattering has also been  studied in other papers in an effort to provide information to fill the phase-diagram of different high-energy scattering regimes. In this energy and impact parameter space different descriptions such as eikonal scattering, Born scattering and string excitations apply in different regions. For a discussion see \cite{Giddings:2009iw} and references therein.

In this paper, we focus on the transplanckian scattering studied in \cite{Giudice:2001ce}. In Section 2 we summarise the calculation of \cite{Giudice:2001ce} for transplanckian scattering at the LHC, presenting the final result for the eikonal amplitude. We then study the signal and background for this process at LHC energies in Section 3. In Section 4, we investigate the extent of the computability region by considering the corrections to the eikonal amplitude. In Section 5, we consider the effect of a possible LHC upgrade to 33~TeV centre-of-mass energy on the signal to background ratio as well as the calculability region and compare with 14~TeV before we conclude.

\section{Scattering amplitude in the eikonal approximation}  
Here we consider graviton exchange in parton-parton scattering in the transplanckian limit. The scattering at transplanckian energies can be described by classical physics in a non-perturbative approach. In this limit, the gravitational field is weak and non-linear graviton couplings can be ignored. In order to tackle the non-perturbative nature of the problem, the eikonal approximation is employed. The approximation is valid when the momentum transfer is small, or equivalently the scattering angle is small. An infinite set of Feynman diagrams is then resummed, and the result is valid when
\begin{equation}
M_D/\sqrt{s}\ll 1 \mbox{ and  } -t/s\ll 1.
\label{limits}
\end{equation}

Details of the calculation are given in \cite{Giudice:2001ce}, where it was found that the eikonal amplitude is
\begin{equation}
\mathcal{A}_{eik}=-2is\int d^2b e^{iq_\perp b}(e^{i\chi}-1),
\end{equation}
with the eikonal phase $\chi$ given by
\begin{equation}
\chi(b)=\frac{1}{2s}\int\frac{d^2q_\perp}{(2\pi)^2}e^{-iq_{\perp}b}\mathcal{A}_{Born}(q_\perp^2)=\bigg(\frac{b_c}{b}\bigg)^n,
\end{equation}
and
\begin{equation}
b_c=\bigg[\frac{(4\pi)^{\frac{n}{2}-1}s\Gamma(n/2)}{2M_D^{n+2}}\bigg]^{1/n}.
\label{bc}
\end{equation}
This final result for the amplitude of the infinite series of ladder and cross-ladder diagrams can be written more simply as
\begin{eqnarray}
\mathcal{A}&=& 4\pi s b^{2}_{c} F_{n}(b_c q),
\label{amplitude}
\end{eqnarray}
where $q\equiv |q_\perp |\simeq \sqrt{-t}$, and the functions $F_n$ are
\begin{equation}
F_n(y)=-i\int^\infty_0 dx\,x\,J_0(xy)(e^{ix^{-n}}-1),
\end{equation} with $J_0$ the zero-th order Bessel function.
These functions are calculated numerically to high precision in our analysis.

\section{Phenomenology of transplanckian scattering}

\subsection{Signal and background at LHC energies}
The amplitude in Eq.~\ref{amplitude} was implemented in a Monte Carlo integration program to calculate the total cross section for this scattering process. For the LHC one should consider all possible combinations of quarks and gluons. The matrix element is the same for all combinations as expected from this classical picture where the spin of the scattered particles does not affect the interaction. The kinematical regions of interest within the eikonal approximation are those of $\sqrt{\hat{s}}/M_D\gg 1$ and $-\hat{t}/\hat{s}\ll 1$.
  This implies that the appropriate regions are those of high rapidity separation between the two jets, $\Delta\eta\gg 1$, where
  \begin{equation}
  \Delta\eta=y_3-y_4=\mbox{ln}(-\hat{s}/\hat{t}-1)=\mbox{ln}\bigg[\frac{1+\mbox{cos}\hat{\theta}}{1-\mbox{cos}\hat{\theta}}\bigg]
  \end{equation} with $\hat{\theta}$ the centre-of-mass scattering angle, and the region of high centre-of-mass energy $M_{JJ}=\sqrt{\hat{s}}\gg M_{D}$. 
  
The differential cross section for partonic scattering is
\begin{equation}
\frac{d\hat{\sigma}}{d\Delta\eta}=\frac{\pi b_c^4\hat{s}e^{\Delta\eta}}{(1+e^{\Delta\eta})^2}|F_n(y)|^2.
\label{difeta}
\end{equation} 
This must be convoluted with the parton distribution functions of the proton to give the hadronic cross section:
\begin{equation}
\frac{d\sigma}{d\Delta\eta}=\int dx_1dx_2 \sum_{i,j} f_i(x_1,Q^2)f_j(x_2,Q^2) \frac{d\hat{\sigma}}{d\Delta\eta}.
\label{hadron}
\end{equation}
 The factorisation scale $Q$ for the parton distribution functions (PDFs) is chosen following the prescription in \cite{Emparan:2001kf}, that is  $Q^2=b_s^{-2}$ if $q>b_c^{-1}$ and $Q^2=q^2$ otherwise, with $b_s=b_c(n/qb_c)^{1/(n+1)}$. The cross sections are calculated with the leading order (LO) PDF set MSTW2008LO \cite{Martin:2009iq}\footnote{We note that in \cite{Giudice:2001ce} the PDF set used is the leading order CTEQ5L \cite{Lai:1994bb} set. We checked that we agree by using CTEQ5L to reproduce some of the plots in \cite{Giudice:2001ce}.}. To ensure we stay in the region of validity of the approximation given above, suitable cuts must be applied to the final-state jets. The cuts used are $p_{T}>$100~GeV, $|\eta|<5$ for detector acceptance and $M_{JJ}>9$~TeV. We note that the detector acceptance cuts $p_{T}>$100~GeV and $|\eta|<5$ are applied in all subsequent plots in this paper.

The background for this scattering process is the dijet cross section from QCD scattering of quarks and gluons. The LO matrix elements squared can be found in \cite{Ellis:1991qj}. The background cross section is also calculated with our VEGAS Monte Carlo integrator with factorisation and renormalisation scales set equal to $p_T/2$. We note that interference effects between the LO QCD background and the signal are negligible, as they are only relevant for a small subset of QCD s-channel scattering diagrams. Interference effects would also come into play for more combinations of partons if we consider the NLO QCD background. 

\begin{figure}
\centering
\includegraphics[scale=0.8]{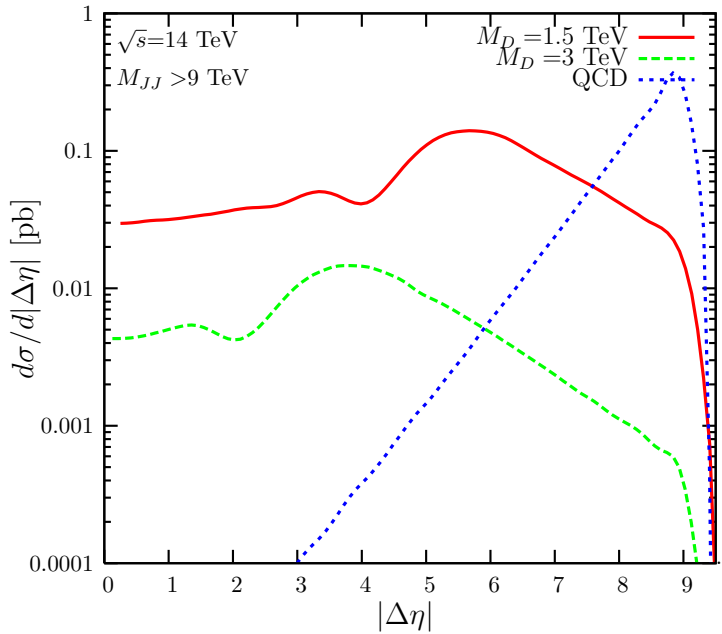}
\caption{Differential cross section for the signal for $n=$6, $M_D$=1.5 and 3~TeV and the QCD background.}
\label{dif}
\end{figure}

\begin{figure}
\centering
\includegraphics[scale=0.8]{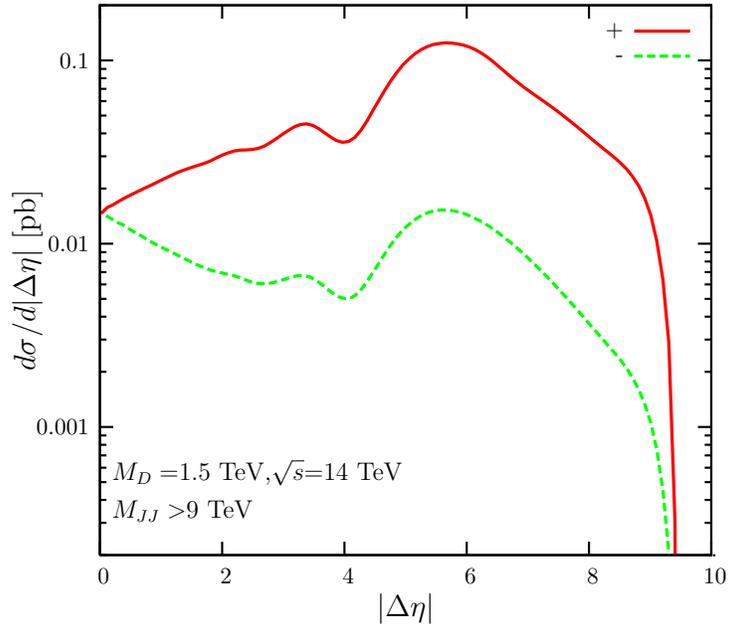}
\caption{Contributions to the differential cross section for $n=$6 and $M_D$=1.5~TeV from positive and negative $\Delta\eta$.}
\label{plusmin}
\end{figure}

 The variable chosen for the plots is $|\Delta\eta|$, as experimentally we cannot distinguish between the two jets. Therefore events with large momentum transfer, where the partons are scattered backwards retracing their paths, cannot be distinguished from small momentum transfer events. Small and negative $\Delta\eta$ regions fall outside the validity of the eikonal approximation but their contribution is expected and checked to be small within the eikonal formalism due to the small values of $F_n(y)$ at large $y=b_cq$ and thus large momentum transfer. The differential cross sections of the signal for $n=6$, for $M_D=1.5$~TeV and 3~TeV and the background are shown in Fig.~\ref{dif} for $\sqrt{s}=14$~TeV at the LHC. For $M_D=1.5$~TeV, we decompose the signal differential cross section $d\sigma/d|\Delta\eta|$ into contributions from positive and negative values of $\Delta\eta$ in Fig.~\ref{plusmin}, which confirms that the contribution included from regions that fall outside the eikonal approximation is small.
\begin{figure}
\centering
\includegraphics[scale=0.8,trim=2cm 0 0 0]{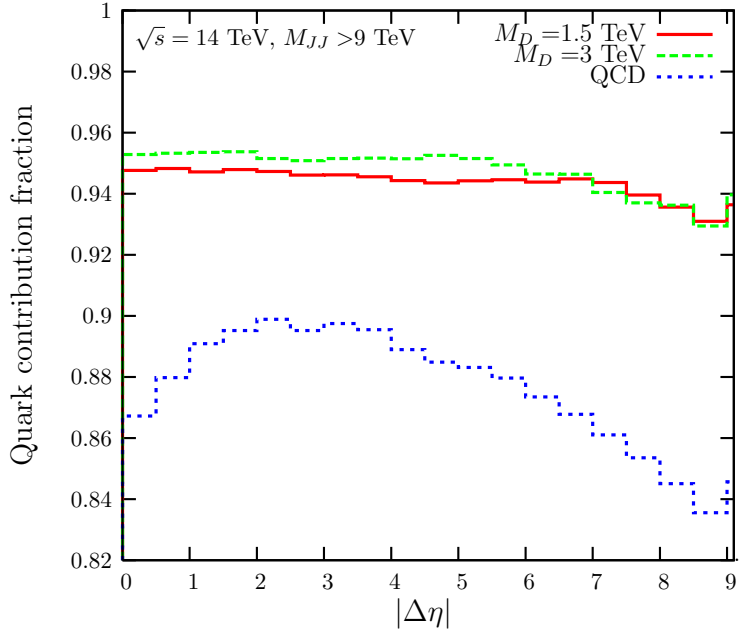}
\caption{Fraction of total cross section contributed by quark scattering for the signal for $n=6$ at $M_D=$ 1.5 and 3~TeV and the background.}
\label{quark}
\end{figure}

Several features emerge from Fig.~\ref{dif}. The fall-off at high rapidity separations exhibited by both the background and the signal is caused by the decrease of the PDF values at high $x$  as the imposed $p_T$ cut forces a large dijet mass which implies large momenta fractions $x$ and correspondingly small PDFs.
For most values of $\Delta\eta$ the values of $x_1,x_2$ remain close to $\sqrt{\hat{s}_{min}/s}$, but at high $\Delta\eta$, $x_1$ and $x_2$ rapidly approach one.
In any case, there is a maximum value of allowed  $\Delta\eta$ from kinematical restrictions. The momenta fractions are given in terms of $\bar{y}=(y_3+y_4)/2$ and $\Delta\eta$ by 
\begin{equation}
x_1=\frac{2p_T}{s}e^{\bar{y}}\mbox{cosh}\frac{\Delta\eta}{2} \mbox{  and  }  x_2=\frac{2 p_T}{s}e^{-\bar{y}}\mbox{cosh}\frac{\Delta\eta}{2}.
\end{equation}
Multiplying the two, we obtain
\begin{equation}
x_1x_2= \frac{4p_T^2}{s^2}\mbox{cosh}^2\frac{\Delta\eta}{2} ,
\end{equation}
which gives a maximum value of $\Delta\eta$=9.88 for $\sqrt{s}$=14~TeV at the LHC and minimum jet transverse momentum $p_T=100$~GeV. We checked that at 14~TeV the imposed cut on the individual rapidities of the jets $|\eta|<$5 does not alter the behaviour of the cross section in the region of high rapidity separation. In this region of high rapidity separation between the jets, the QCD background receives significant contributions from BFKL dynamics \cite{Mueller:1986ey,Orr:1997im}, which are not taken into account here. 

We also note that for this high $M_{JJ}$ cut, the values of $x_1$ and $x_2$ are high and the dominant contribution to the cross section comes from quark-quark scattering. This is shown in Fig.~\ref{quark} where we plot the fraction of the total cross section we obtain when we turn off the gluon contributions. The quark contribution for the signal is larger as all combinations of partons have equal weight within the eikonal approximation, while for the QCD cross section different colour factors change the weight of each contribution, i.e. the effective PDF is $g+\frac{4}{9}\sum (q+\bar{q})$ for the QCD cross section \cite{Ellis:1991qj}, compared to $g+\sum (q+\bar{q})$ for the signal. The values of $x_1$ and $x_2$ probed are roughly constant in the range $|\Delta\eta|<9$. The difference between the two values of $M_D$ is caused by the $M_D$ dependence of the PDF scale. For the QCD background in Fig.~\ref{quark}, the variation of the quark scattering fraction comes from the $|\Delta\eta|$ dependence of the renormalisation and factorisation scales.

Another effect evident in Fig.~\ref{dif} is that the differential cross section is larger for smaller $M_D$. From Eqs.~\ref{bc} and \ref{difeta}, we know that the differential cross section is proportional to $M_D^{-4(1+2/n)}$. The cross section also depends on $M_D$ through the $y=b_cq$ argument of the functions $F_n(y)$. Large $M_D$ and thus small $b_c$ tend to increase the value of $F_n$ and thus of the cross section, but for $n=6$ this has a $\emph{smaller}$ effect than the $M_D^{-16/3}$ factor from $b_c^4$ in the expression for the cross section.
  
\begin{figure}
\centering
\includegraphics[scale=0.8]{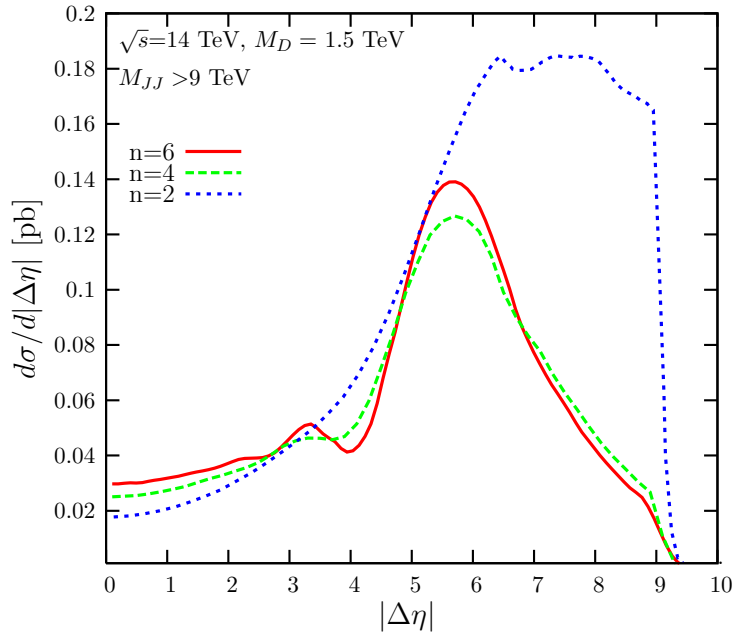}
\caption{Differential cross section for $n=$2, 4, 6 and $M_D$=1.5~TeV.}
\label{ndep}
\end{figure}

The peak structure observed in the differential cross section is due to the form of the functions $F_n(y)$ and corresponds to a diffraction pattern of the scattered particles. From
\begin{equation}
\frac{d\hat{\sigma}}{d\Delta\eta}=\frac{\pi b_c^4\hat{s}e^{\Delta\eta}}{(1+e^{\Delta\eta})^2}|F_n(y)|^2,
\end{equation}
the maxima in the partonic cross section are obtained from
\begin{equation}
-\frac{y}{|F_n(y)|}\frac{d|F_n(y)|^2}{dy}=1-e^{-\Delta\eta},
\label{maxima}
\end{equation} 
as $y=b_c\sqrt{\hat{s}}/\sqrt{1+e^{\Delta\eta}}$. This cannot be solved exactly but the solutions can be estimated numerically to locate the peaks, if those are at large $\Delta\eta$, so that one can ignore the second term in the right-hand side of Eq.~\ref{maxima}. The number of extra dimension changes the structure of the functions $F_n$. For comparison, in Fig.~\ref{ndep} we show the differential cross section for $n=$2, 4, 6 on a linear scale at 14~TeV with the same cuts as in Fig.~\ref{dif}. The different position of the peaks for each $n$ could, at least in principle,  be used to experimentally extract the number of extra dimensions. 
\subsection{Comparison of eikonal signal to other signals} 
\begin{figure}
\centering
\includegraphics[scale=0.8]{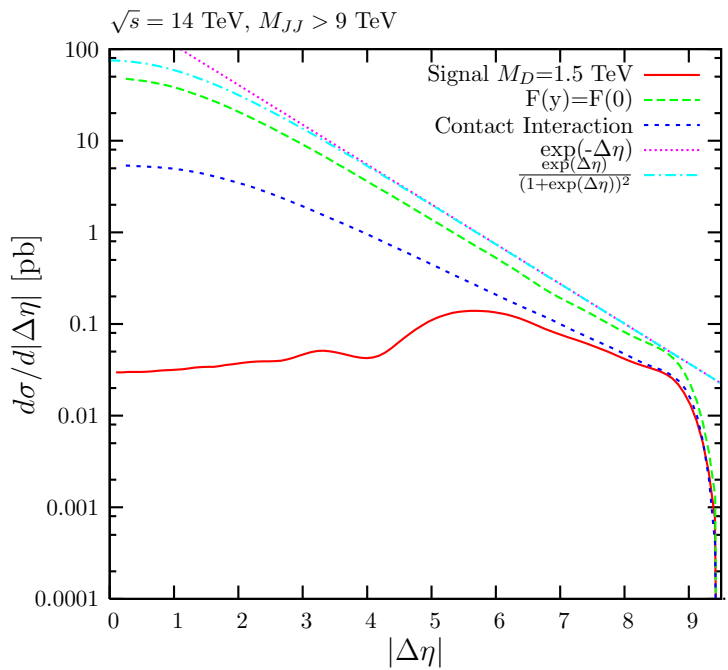}
\caption{Differential cross section for different choices of interaction. The contact interaction constant has been adjusted to be of the same order of magnitude in the matching behaviour region, as are the coefficients of the two functions.}
\label{contact}
\end{figure}

In order to better understand the shape of the cross section, we consider a test case where $F_6(y)$ is fixed at its $y=0$ value, i.e. its value at large rapidity separations. In this case, the cross section is expected to keep increasing monotonically on moving from large to small rapidity separations, giving a straight line at sufficiently large $\Delta\eta$ in a log plot. This is explained simply by Eq.~\ref{difeta}, which at large $\Delta\eta$ simplifies to
\begin{equation}
\frac{d\hat{\sigma}}{d\Delta\eta}=\pi b_c^4\hat{s}e^{-\Delta\eta}|F_n(y)|^2.
\label{constant}
\end{equation}
Note that similar behaviour is expected for a four-particle contact interaction such as the one proposed in \cite{Eichten:1984eu}, where the contact interaction for left-handed quarks has the form
\begin{equation}
\mathcal{L}=\frac{n_0 g^2}{2\Lambda^2}\bar{q}_L\gamma^{\mu}q_L\bar{q}_L\gamma_{\mu}q_L \mbox{ with }\,\, n_0=\pm 1.
\label{contL}
\end{equation}
The amplitudes squared are given in \cite{Eichten:1984eu}. For comparison, we show in Fig.~\ref{contact}  the cross-section results for the signal, for fixed $F_6(y)=F_6(0)$, for the contact interaction and for the functions  $e^{-\Delta\eta}$ and $e^{\Delta\eta}/(1+e^{\Delta\eta})^2$ to illustrate the different functional behaviours. 
The contact interaction constant $g^2/\Lambda^2$ and the coefficients of the two functions are adjusted so that their values match those of the signal at the matching region of large $\Delta\eta$. 

Of course, there are strong experimental limits on the contact interaction strength from several experiments, see for example the summary table in \cite{Amsler:2008zzb} and the recent results from the D0 collaboration \cite{:2009mh}. Recently, the first LHC results were presented  by the ATLAS~\cite{Collaboration:2010eza} and CMS~\cite{Khachatryan:2010te} collaborations. The two collaborations searched for deviations from the SM prediction for the angular distribution of dijet events and raised the 95\% c.l.\ exclusion limit to 3.4 and 4~TeV respectively. 

Here, we are only interested in establishing the functional dependence of the differential cross section on $\Delta\eta$ for the different cases listed above. We note that the behaviour described by Eqs.~\ref{difeta}, \ref{constant} refers to the partonic cross section, but it is also shown by the hadronic cross section for all but very high $\Delta\eta$. This can again be explained by the high $M_{JJ}$ cut and the fact that the cross section is a rapidly decreasing function of $M_{JJ}$. This leads to approximately constant values of $x_1$ and $x_2$ and thus $\hat{s}$, which do not allow the integration in Eq.~\ref{hadron} to alter the functional dependence of the cross section on $\Delta\eta$.

We also note that the differential cross section is symmetric in $\Delta\eta$ for the contact interaction and for the case of fixed $F(y)$, which leads to a factor of two difference at the matching region between the full treatment and this simplified test case of constant $F$. The graviton signal cross section is not symmetric in $\Delta\eta$, as the value of $y$ increases with decreasing $\Delta\eta$ and $F_n(y)\rightarrow 0$ as $y\rightarrow \infty$. On the other hand, the QCD background $\emph{is}$ symmetric in $\Delta\eta$, assuming a renormalisation and factorisation scale choice symmetric in $\Delta\eta$. 

\begin{figure}
\centering
\includegraphics[scale=0.8]{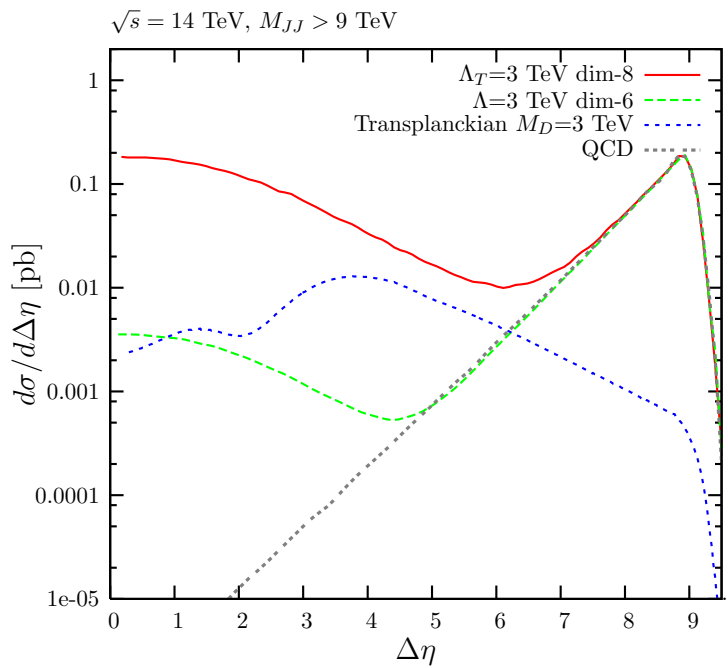}
\caption{Differential cross section for interactions originating from dimension-8 operators in comparison with dimension-6 operators, the transplanckian signal for $n=6$ and the QCD background.}
\label{dim8}
\end{figure}

In addition to the dimension-6 operators of Eq.~\ref{contL}, which arise in BSM theories such as technicolour, but also in extra dimensional theories from one-loop graviton exchange~\cite{Giudice:2003tu}, one can also compare the transplanckian signal to the dimension-8 operators that arise in extra-dimensional theories. In this case, the amplitude for a single virtual graviton exchange is represented by the energy momentum tensor squared and the UV divergence is regularised arbitrarily by introducing the coupling constant $\Lambda_T^{-4}$. This is used to calculate the virtual graviton effects in the cisplanckian region. The value of the constant is expected to be nearly the value of $M_D$ but precise numerics are unknowable. The corresponding matrix elements for dijet production have been calculated in \cite{Atwood:1999qd}. The differential cross section $d\sigma/d\Delta\eta$ for the signal is added to the background for the dimension-6 and 8 operators and shown in Fig.~\ref{dim8} for $\Lambda_T$=3~TeV.
We note that we choose the value of the fundamental scale $M_D$, the strength of the contact interaction $\Lambda$ and the UV cut-off $\Lambda_T$ to be the same for illustration purposes. One can appropriately rescale the cross sections for the dimension-6 and dimension-8 operators, as in the region where the signal dominates over the background, they have a simple dependence on the coupling constant: $\Lambda^{-4}$ and  $\Lambda_T^{-8}$ respectively. 

The shape of the signal can evidently be used to distinguish between the different types of interactions with the eikonal amplitude giving a distinctive peaking structure. We emphasize here that the calculation for the dimension-6 and dimension-8 operators signals is strictly not valid in this region of high centre-of-mass energy larger than the cut-off scales. The purpose of Fig.~\ref{dim8} is simply to show the functional behaviour of the cross section for the chosen region of $\Delta\eta$ for different operators. 

\subsection{PDF and PDF scale dependence of the cross sections}
The dependence of the results on different renormalisation and factorisation scale choices has also been investigated. The scale affects the signal through the PDF values and the background through both the PDF values and the explicit value of $\alpha_s$ in the matrix elements, as we have set the renormalisation scale equal to the factorisation scale in this study. The relations connecting $\Delta\eta$ and possible scale options are
\begin{eqnarray}
 q&=&\frac{\sqrt{\hat{s}}}{\sqrt{1+e^{\Delta\eta}}},\\
 p_T&=&\frac{\sqrt{\hat{s}}e^{\Delta\eta/2}}{1+e^{\Delta\eta}},\\
 b_s^{-1}&=&b_c^{-1}\bigg(\frac{n}{qb_c}\bigg)^{-1/(n+1)},
\end{eqnarray}with $b_c$ given in Eq.~\ref{bc}.

\begin{figure}
\centering
\includegraphics[scale=0.8]{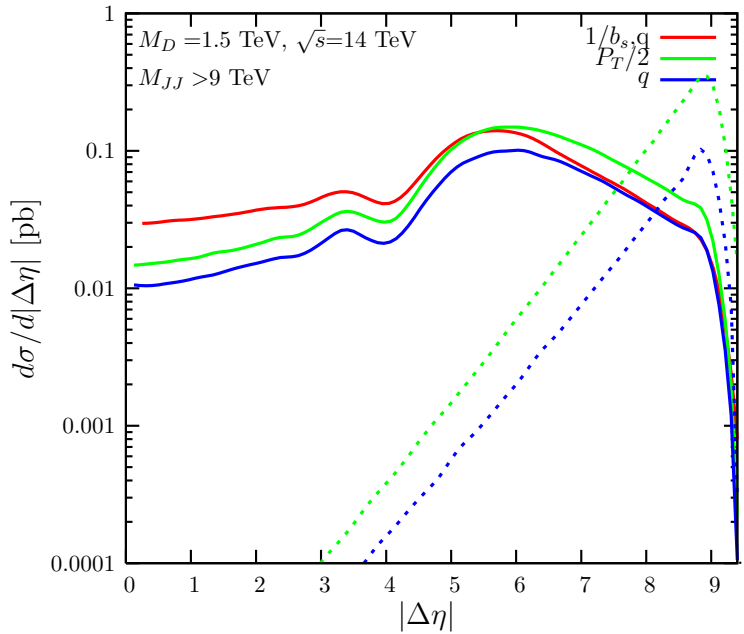}
\caption{Dependence of the cross section on scale choices for both the background and the signal for $n=6$.}
\label{scalecr}
\end{figure}

The effect of the scale choice on the LO differential cross sections is shown in Fig.~\ref{scalecr}. 
The choice $Q=q=\sqrt{-t}$ matches the prescription in \cite{Giudice:2001ce} for $\Delta\eta$ larger than $\sim$7, in agreement with the crossing between $q$ and $1/b_c$. In general, we see that smaller scale choices lead to larger cross sections. One should of course consider next-to-leading order calculations to reduce this dependence on the scale choice, as has been pursued in \cite{Lodone:2009qe}.  

\begin{figure}
\begin{minipage}[b]{0.5\linewidth} 
\centering
\includegraphics[trim=0.7cm 0 0 0,scale=0.56]{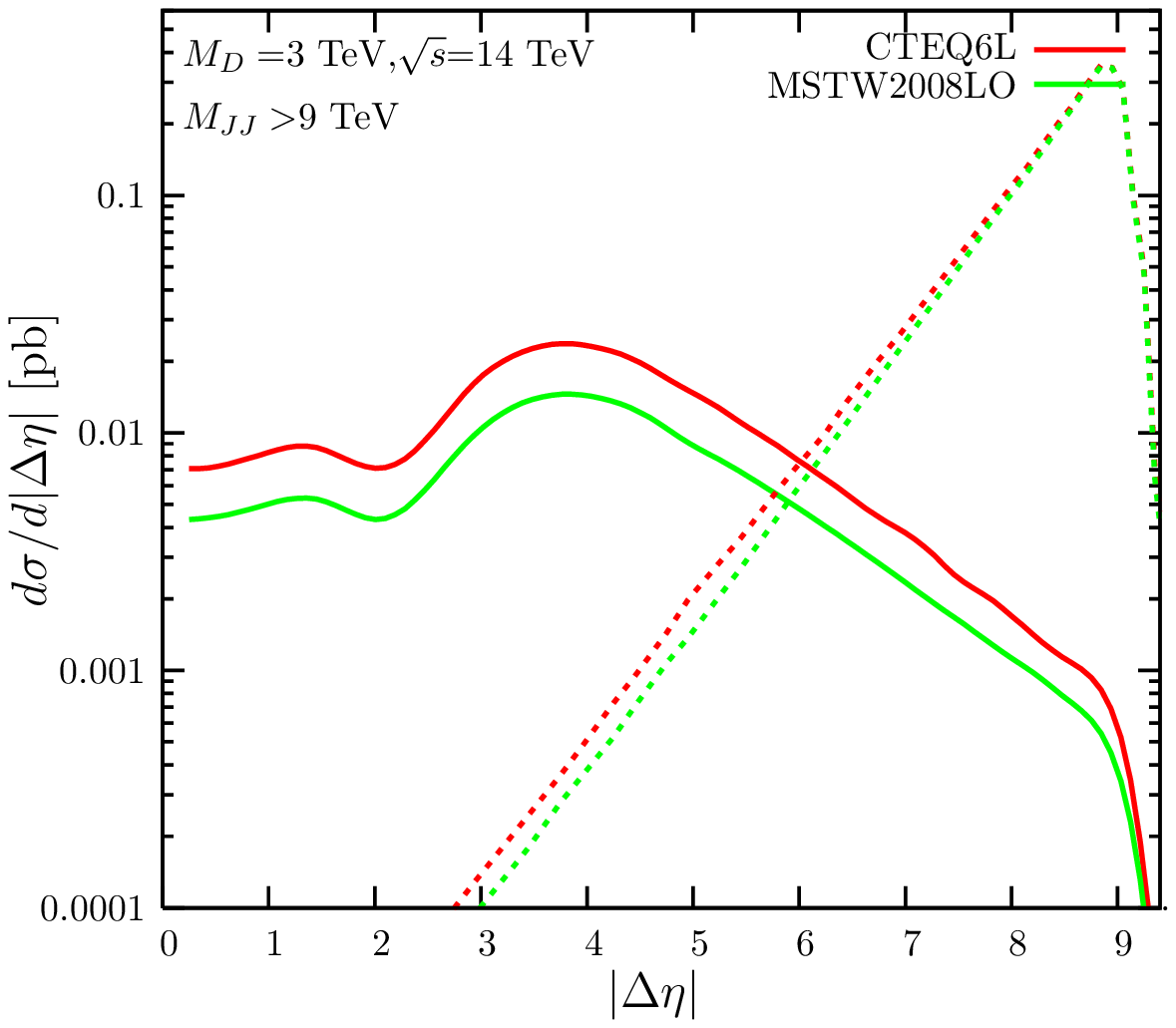}
\caption{Dependence of the cross section on PDF choices for both the $n=6$ signal and the background.}
\label{pdf2}
\end{minipage}
\hspace{0.3cm} 
\begin{minipage}[b]{0.5\linewidth}
\centering
\includegraphics[trim=1.2cm 0 0 0,scale=0.56]{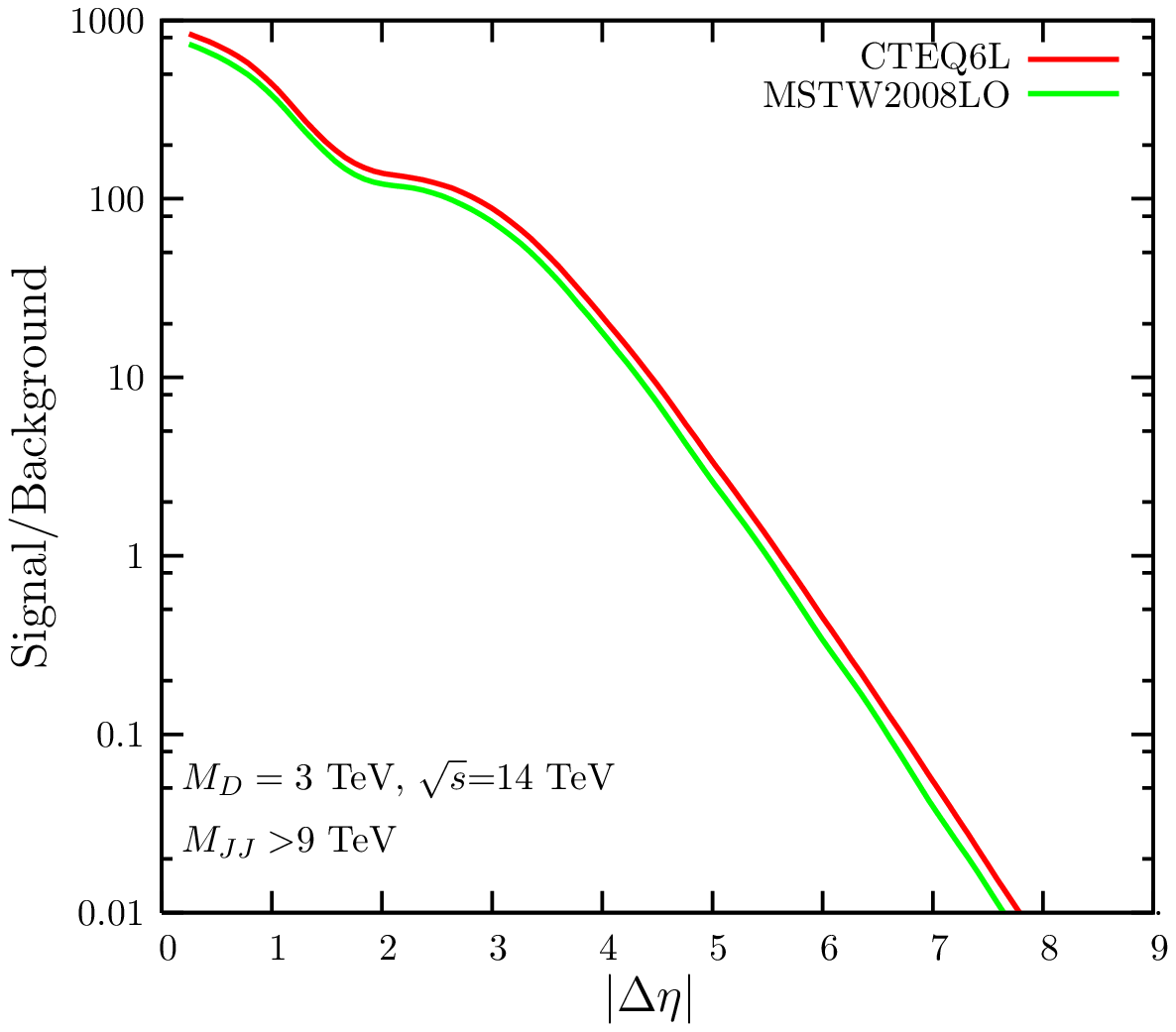}
\caption{Dependence of the signal to background ratio on the choice of PDF set, for the signal of $n=6$.}
\label{pdf}
\end{minipage}
\end{figure}

As an additional check, we also consider two different PDF sets. The differential cross section for the signal at $M_D=$ 3~TeV and the background is shown for two different leading order PDF choices, MSTW2008LO~\cite{Martin:2009iq} and CTEQ6L \cite{Pumplin:2002vw}, in Fig.~\ref{pdf2}. There is a difference of about $35\%$ between the two sets for the signal, while the difference is smaller for the background. For the signal the difference is mostly caused by the difference of the PDF values for the valence $u$ and $d$ quarks which at $x=0.7$ differ in the two sets by almost 20$\%$ and 40$\%$ respectively at $Q=100$ GeV. For the background, this difference is partially compensated to some extent by the value of $\alpha_s$, for which the values of the two sets differ in the opposite way by about 20$\%$ at $M_Z$. In Fig.~\ref{pdf}, we show the signal to background ratio for the two PDF sets. The difference in the signal is partially compensated by the difference in the background for small $|\Delta\eta|$ values.
\subsection{Total cross section for eikonal signal}
\begin{figure}
\centering
\includegraphics[scale=0.8]{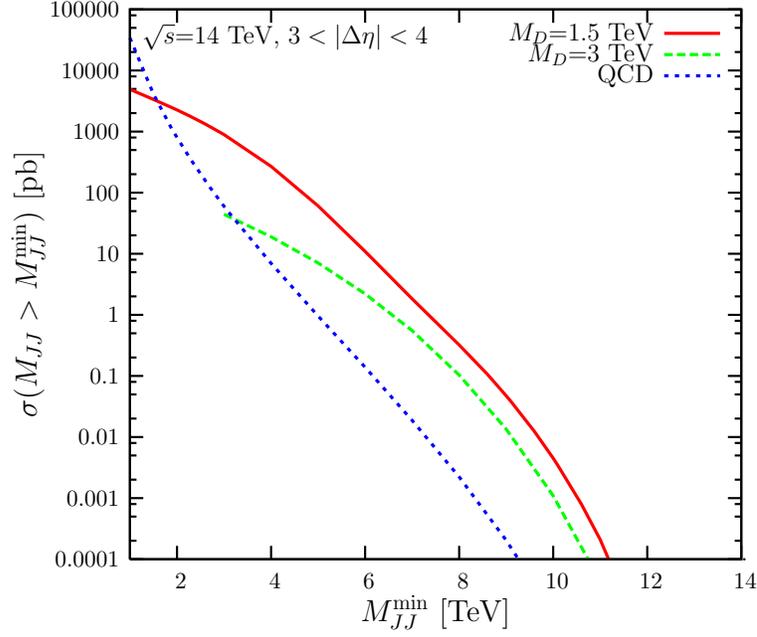}
\caption{Total dijet cross section as a function of the $M_{JJ}$ cut for the signal at $M_D=$1.5 and 3~TeV, for $n=6$ and the background.}
\label{figure6}
\end{figure}

In Fig.~\ref{figure6}, the total cross section is shown as a function of the $M_{JJ}$ cut for both the signal and the background. The rapidity separation is forced to be in the range  $3<|\Delta\eta|<4$, and the $p_T>100$~GeV and $|\eta|<5$ cuts remain in place. The choice of the acceptance region of $|\Delta\eta|$ is a compromise between staying in the kinematical region allowed within the eikonal approximation (large $\Delta\eta$) and maximising the signal to background ratio (small $\Delta\eta$). We note that the background falls more rapidly than the signal with increasing $M_{JJ}^{min}$. Therefore the search for signal against background will be easier for a higher cut provided the event rate remains high enough for detection. We also note that the cross section is smaller for larger $M_D$, as it is suppressed by inverse powers of $M_D$. Moreover the cut needed to ensure we remain in the transplanckian region increases with larger $M_D$, leading to smaller cross sections and lower event rates. We thus expect better prospects of discovery for smaller $M_D$ values. 

\begin{figure}
\begin{minipage}[b]{0.5\linewidth} 
\centering
\includegraphics[trim=1.3cm 0 0 0,width=7.6cm]{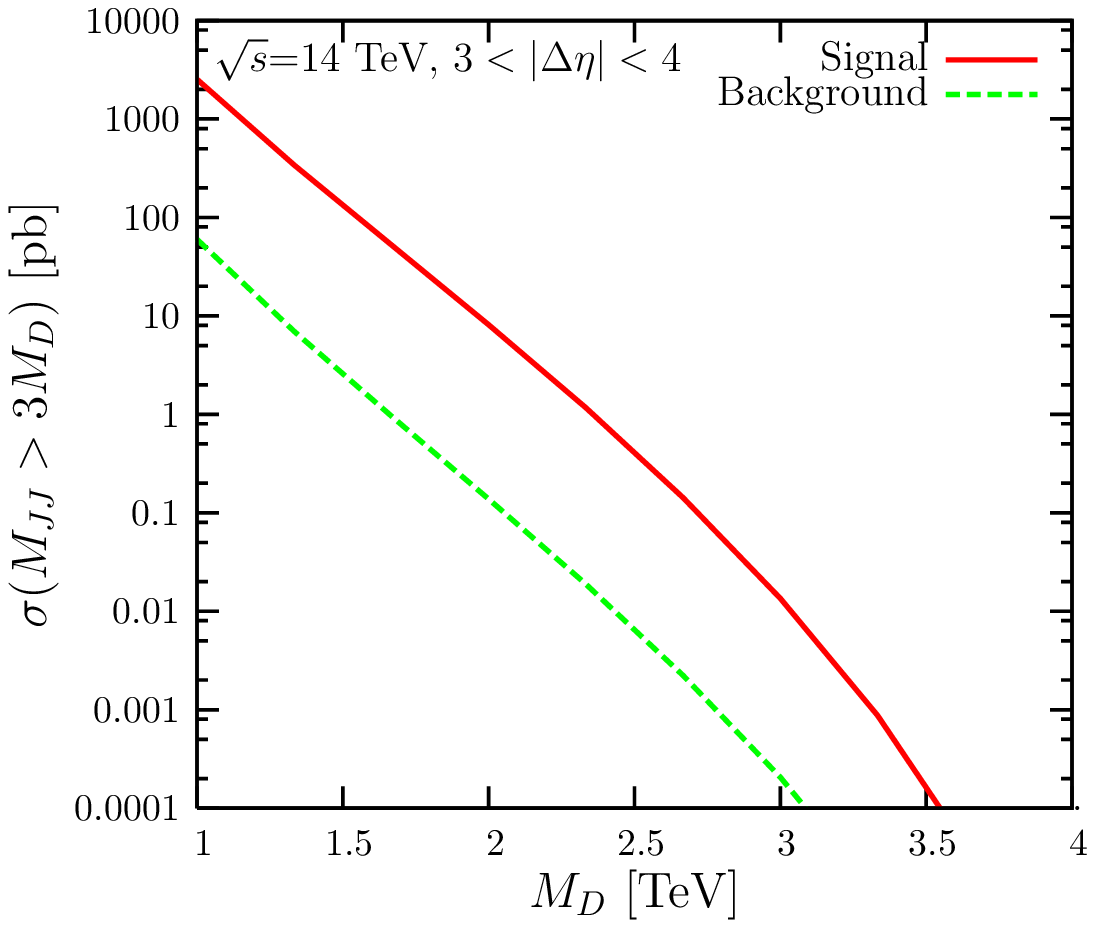}
\caption{Total cross section for $n=6$, as a function of $M_D$ for $M_{JJ}^{min}=3M_D$.}
\label{fig3MD}
\end{minipage}
\hspace{0.3cm} 
\begin{minipage}[b]{0.5\linewidth}
\centering
\includegraphics[trim=1.3cm 0 0 0,width=7.6cm]{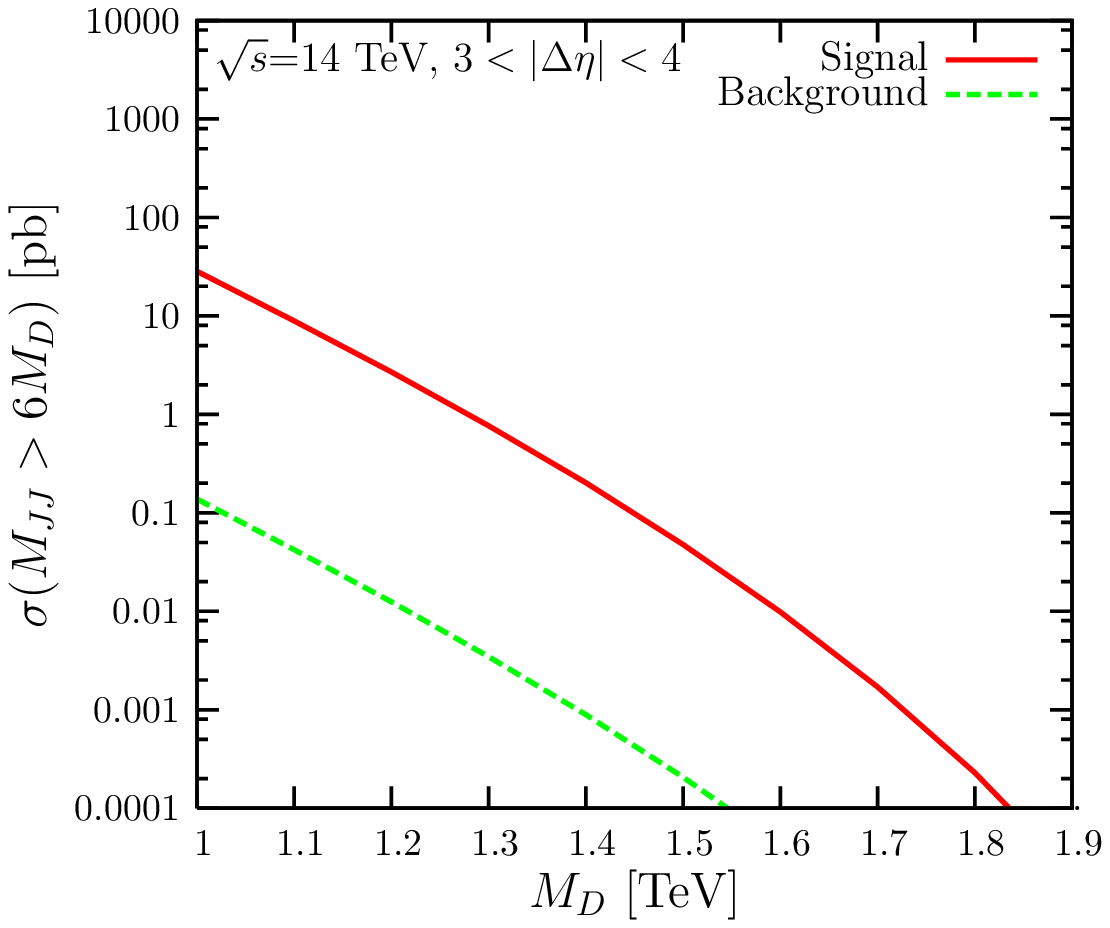}
\caption{Total cross section for $n=6$, as a function of $M_D$ for $M_{JJ}^{min}=6M_D$.}
\label{fig6MD}
\end{minipage}
\end{figure}

Finally, in Figs.~\ref{fig3MD} and \ref{fig6MD}, we show the dependence of the total cross section on the fundamental scale $M_D$, fixing the $M_{JJ}$ cut at 3$M_D$ and the more conservative $6M_D$ respectively, again for $3<|\Delta\eta|<4$. Given the current experimental limits on the fundamental scale $M_D$ with the centre-of-mass energy of 14~TeV available at the LHC, we can barely reach the transplanckian region for the more conservative $M_{JJ}$=6$M_D$ choice, with the PDF values suppressing the cross sections.  Although the more reliably calculated limits from graviton emission are still well under 2 TeV, the indications from dimension-8 and dimension-6 induced operators, such as those discussed in~\cite{Giudice:2003tu} are that $M_D$ may need to be higher, of order several TeV. Although such estimates are uncertain and dependant on the yet-unknown underlying theory of quantum gravity, if correct they would put the transplanckian region nearly out of reach for 14 TeV LHC.  However, the recently discussed higher energy LHC of 33 TeV could still reach transplanckian validity in that case.   That will be the focus of the penultimate section, but first we discuss how to quantify roughly the calculability of the transplanckian signal in the next section.
 

\section{Computability region of transplanckian scattering signal}
The conditions for the calculation to be valid are given in Eq.~\ref{limits}. These relate to the constraint on the energy to be in the transplanckian region and the small scattering angle for the eikonal approximation to be valid. In the transplanckian region scattering can be described by classical physics and is thus only calculable by non-perturbative methods. Imposing also the constraint of large impact parameter, so that the gravitational field is weak, allows us to use the eikonal approximation to calculate the scattering amplitude. The eikonal approximation is only valid for small scattering angles which imposes the second condition. Corrections to the eikonal amplitude originate from both classical and quantum-mechanical effects. The classical effects originate from neglecting terms of order $-t/s$ and $R_S/b$ which are assumed to be very small. The emergence of the terms is discussed below.

For a collision with centre-of-mass energy $\sqrt{s}$, the Schwarzschild radius of the system is formed as:
\begin{equation}
R_S=\frac{1}{\sqrt{s}}\bigg[\frac{8\Gamma(\frac{n+3}{2})}{(n+2)}\bigg](G_D\sqrt{s})^{1/(n+1)},
\end{equation}
with $G_D$ the D-dimensional Newton's constant. In the region where $\sqrt{s}\gg q \gg b_c^{-1}$, $b<b_c$ and the eikonal phase is large, and we are in the classical region. By dimensional analysis the scattering angle of a collision is related to the impact parameter $b$ and the Schwarzschild radius by $\theta\sim (R_S/b)^{n+1}$. Therefore the expansion parameter $-t/s$ is related to the impact parameter of the collision (ignoring factors of order one) by
\begin{eqnarray}
-\frac{t}{s}\sim \frac{G^2_Ds}{b^{2n+2}}\sim \bigg(\frac{R_S}{b}\bigg)^{2n+2}. 
\label{impact}
\end{eqnarray}
Therefore in this region the two conditions coincide. A small $-t/s$ also ensures that the impact parameter is larger than the Schwarzschild radius so that the gravitational field is weak and the risk of black hole formation is avoided. By including non-linear effects Eq.~\ref{impact} will be modified by a factor of $1+\mathcal{O}(t/s)$. 

The other aspect that needs to be controlled for this semi-classical calculation to be valid is the quantum nature of the particles. When $q\ll b_c^{-1}$, $b\sim b_c$, the eikonal phase is of the order one and quantum effects are important. In this region the size of the correction for small $t$ is \cite{Giudice:2001ce}:
\begin{equation}
\frac{1}{sb_c^2}\sim \bigg(\frac{R_S}{b_c}\bigg)^{2n+2}\sim \bigg(\frac{M^2_D}{s}\bigg)^{1+\frac{2}{n}}.  
\end{equation}
Combining the two, the leading classical gravity corrections to the eikonal amplitude are
\begin{equation}
 \mathcal{O}\bigg(-\frac{t}{s}\bigg)+\mathcal{O}\bigg[\bigg(\frac{M_D^2}{s}\bigg)^{1+\frac{2}{n}}\bigg].
\end{equation}
In addition to these, there are quantum gravity corrections to the eikonal amplitude. These may be model-dependent string corrections which we do not consider here. 

\begin{figure}
\centering
\includegraphics[scale=1.0]{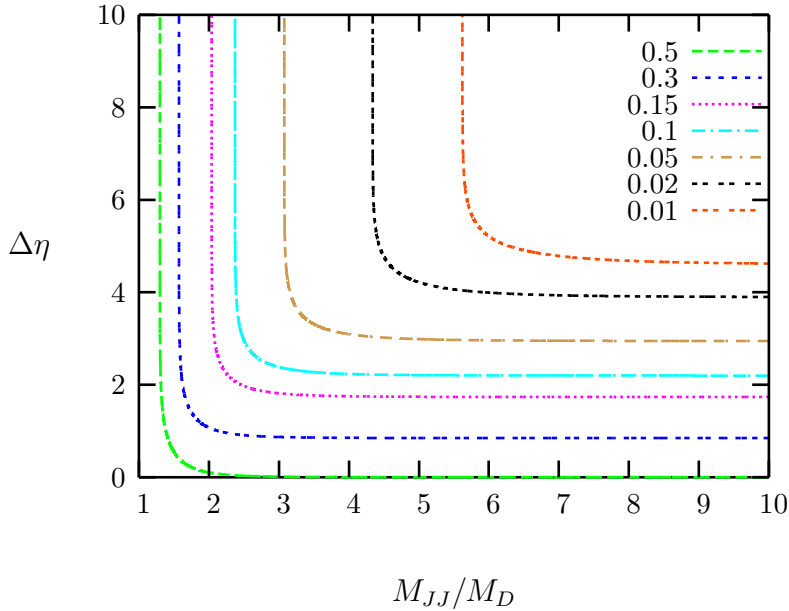}
\caption{Dependence of $\epsilon$ on $\Delta\eta$ and $M_{JJ}/M_D$ for $n=$6. The lines are contours of constant $\epsilon$. }
\label{epsilon}
\end{figure}

In order to visualise and quantify the computability of the two-to-two scattering, we introduce the parameter $\epsilon$, defined as
\begin{equation}
\epsilon=\bigg|\frac{t}{s}\bigg|+\bigg(\frac{M_D^2}{s}\bigg)^{1+\frac{2}{n}}.
\label{epsiloneq}
\end{equation} For any assumed or given value of $M_D$ the parameter $\epsilon$ is well determined from kinematical quantities we used in the previous sections to study the two-to-two scattering, as there is direct correspondence between $-t/s$ and the rapidity separation between the two jets $\Delta\eta$ through
\begin{equation}
-\frac{t}{s}=\frac{1}{1+e^{\Delta\eta}}. 
\end{equation}
Moreover, $\sqrt{s}$ is given by the dijet mass. The dependence of $\epsilon$ on $s$ and $\Delta\eta$ for $n=6$ is shown in the contour plot of Fig.~\ref{epsilon}. The lines are contours of constant $\epsilon$ with the value for each colour line given in the legend. From this contour plot we can determine the region in the $M_{JJ}$ and $\Delta\eta$ space where the calculation is reliable. Computability requires a small value of $\epsilon$ which corresponds to large $\Delta\eta$ and large $M_{JJ}/M_D$.
In the following section, we will use $\epsilon$ to visualise the extent of the calculability region of the theory. 
\section{Comparison of HE LHC and LHC}
\begin{figure}
\centering
\includegraphics[scale=0.8]{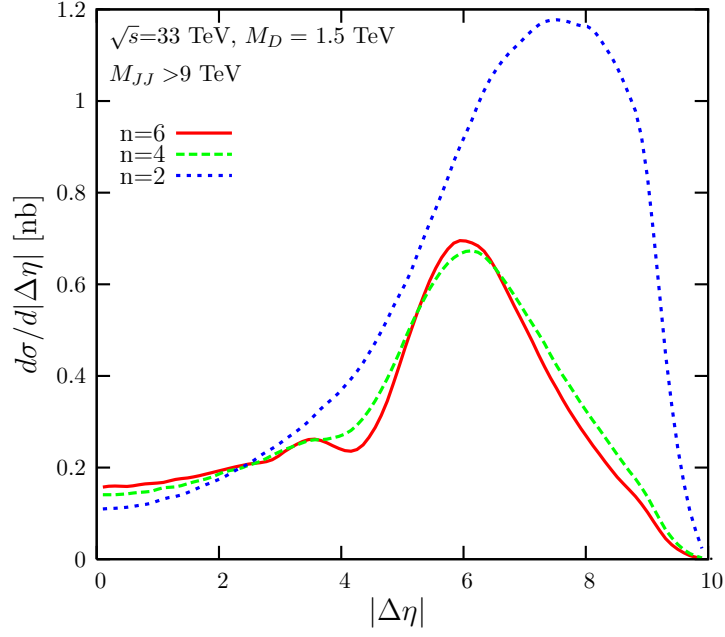}
\caption{Differential cross section $d\sigma/d|\Delta\eta|$ at $\sqrt{s}=$33~TeV for $M_{D}=$1.5~TeV for different n. }
\label{ndep33}
\end{figure}

\begin{figure}
\centering
\includegraphics[scale=0.8]{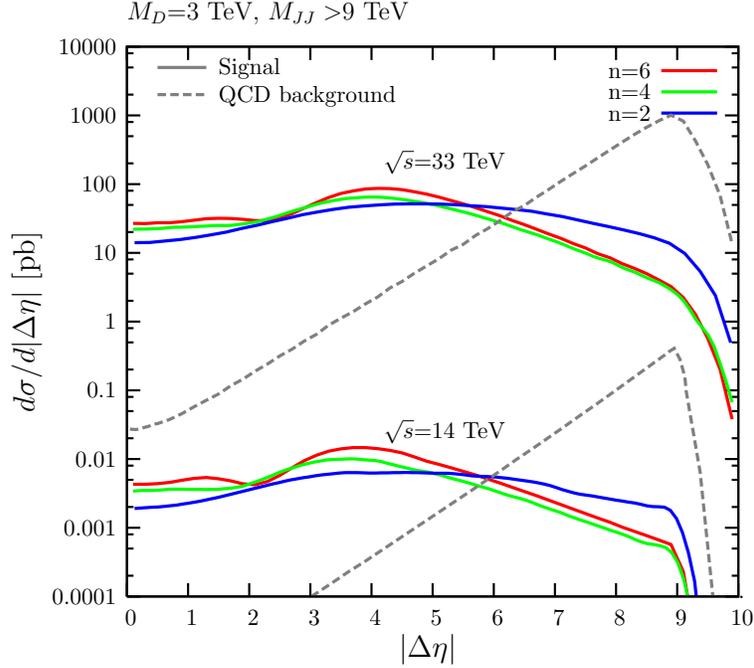}
\caption{Differential cross section $d\sigma/d|\Delta\eta|$ at $\sqrt{s}=$14 and 33~TeV  for $M_{D}=$3~TeV for different n.}
\label{alln}
\end{figure}

Increasing the centre-of-mass energy of the collider would enable us to probe collisions with partonic centre-of-mass energy much greater than $M_D$ and thus well into the transplanckian regime. The proposed energy upgrade for the LHC (HE LHC) based on 20~T magnets, would increase the beam energy to 16.5~TeV \cite{CERN}.  For the HE LHC, the higher centre-of-mass energy of 33~TeV allows us to set the $M_{JJ}$ cut higher while at the same time the total rate for the signal is increased significantly.
For comparison with Fig.~\ref{ndep} we show the corresponding plot for $\sqrt{s}$= 33~TeV in Fig.~\ref{ndep33}. We note that in this plot the cross section units are nb compared to pb we had in Fig.~\ref{ndep}.  
Adding all cases in one plot to better visualise the different effects present, we show in Fig.~\ref{alln} the differential cross section for $M_D=$ 3~TeV and n=2, 4, 6 with the same $M_{JJ}$ cut. While the signal to background ratio is similar for both centre-of-mass energies, the cross sections are four orders of magnitude larger for $\sqrt{s}=$33~TeV leading to more easily measurable event rates.

\begin{figure}
\centering
\includegraphics[scale=0.8]{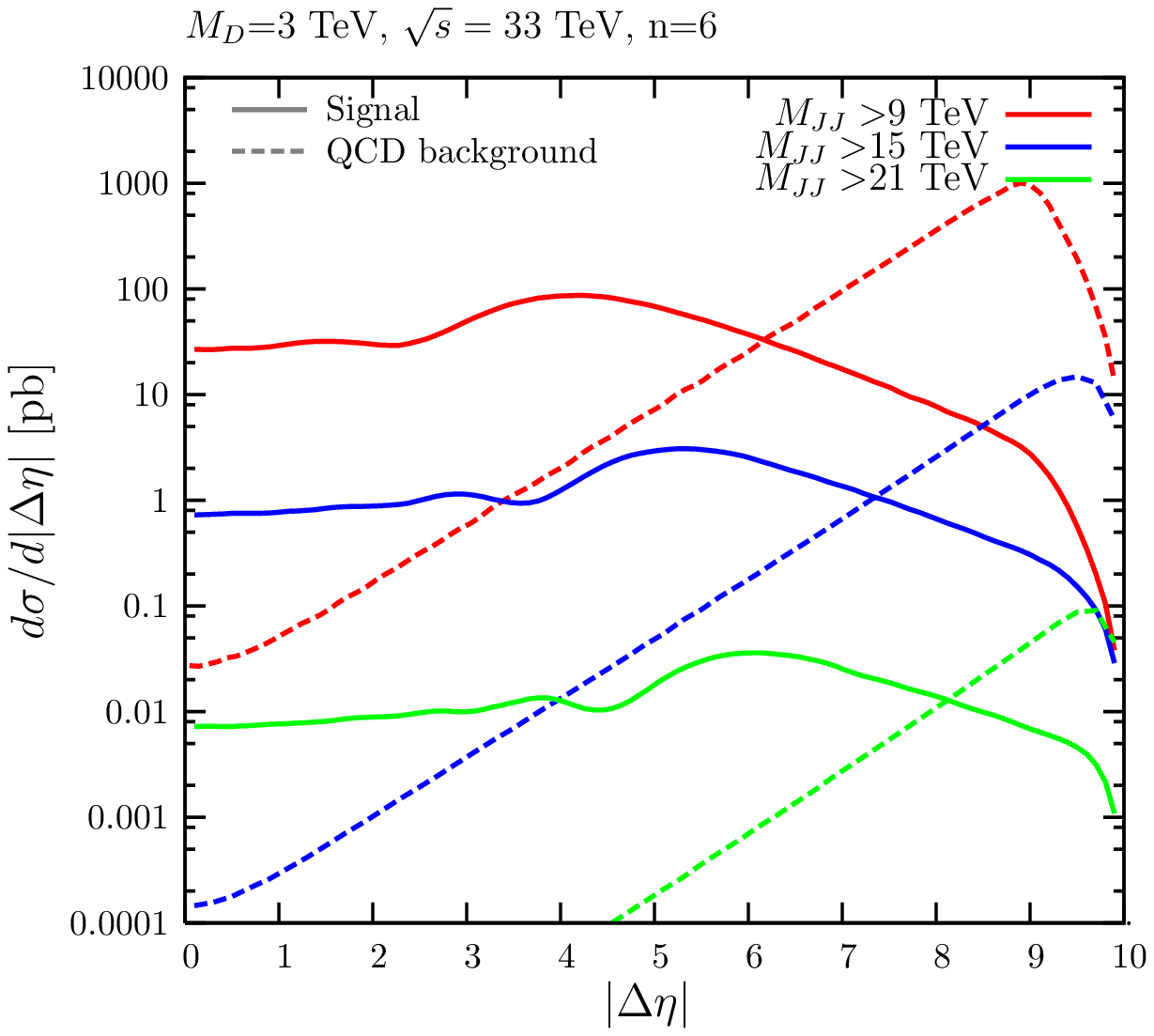}
\caption{Differential cross section $d\sigma/d|\Delta\eta|$ at $\sqrt{s}=$33~TeV  for $M_{D}=$3~TeV and n=6 varying the $M_{JJ}$ cut.}
\label{mjjcut}
\end{figure}

Moreover, we now have the chance to increase the $M_{JJ}$ cut retaining measurable cross sections. The effect of increasing the dijet mass cut is shown in Fig.~\ref{mjjcut}. On increasing the dijet mass cut the crossing point of signal and background moves to the right, and therefore the region where the signal is larger than the background is extended to higher $\Delta\eta$ values. The importance of this is explored in the rest of the section where we examine the effect of increasing the centre-of-mass energy on the calculability region. 

\begin{figure}
\begin{minipage}[b]{0.5\linewidth} 
\includegraphics[trim=1cm 0 0 0,width=7.3cm,height=7cm]{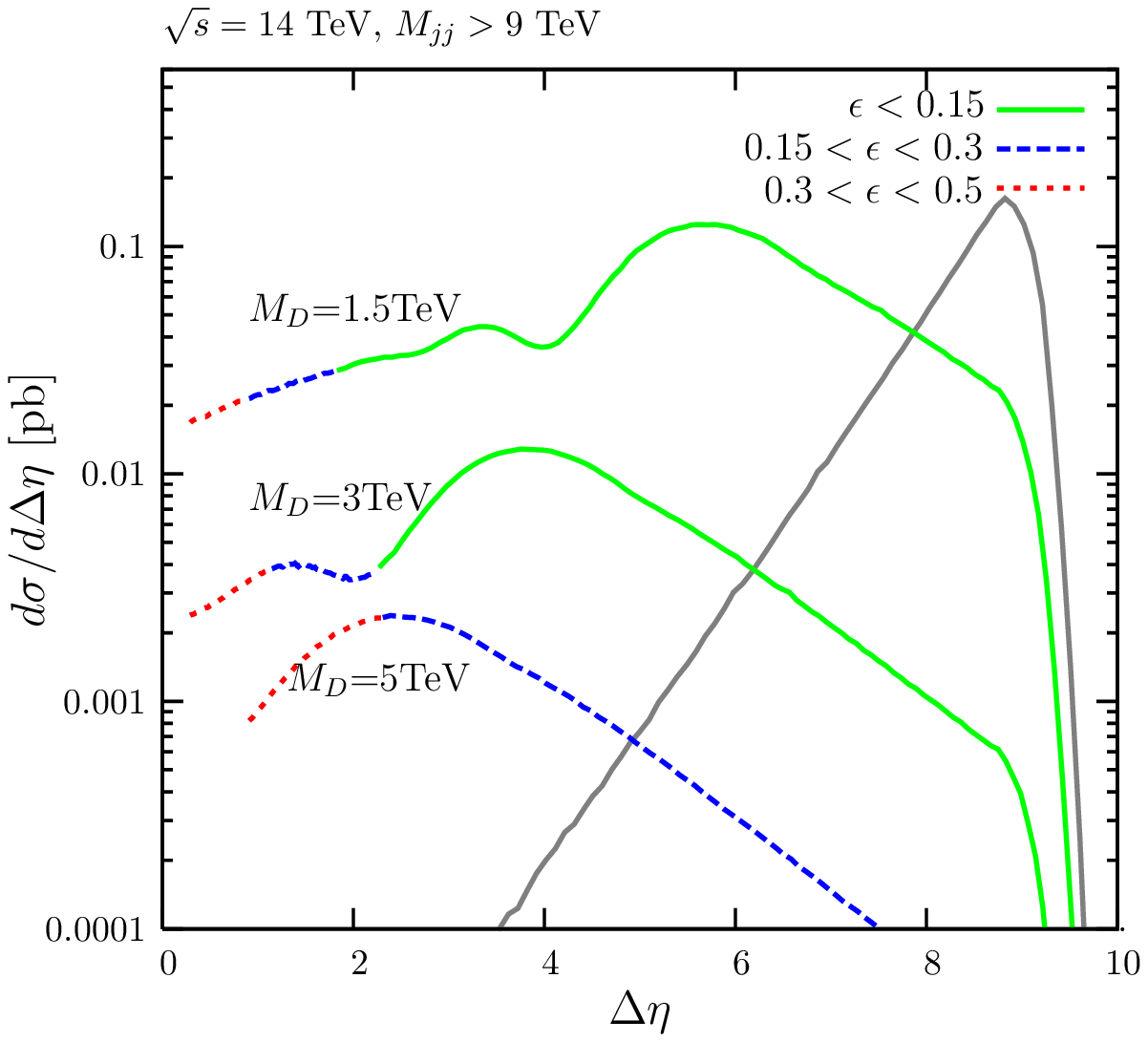}
\caption{Differential cross section $d\sigma/d\Delta\eta$ for $n=6$, at $\sqrt{s}=$14~TeV and $M_{JJ}^{min}=$9~TeV.}
\label{sb14}
\end{minipage}
\hspace{0.2cm} 
\begin{minipage}[b]{0.5\linewidth}
\includegraphics[trim=1cm 0 0 0,width=7.3cm,height=7cm]{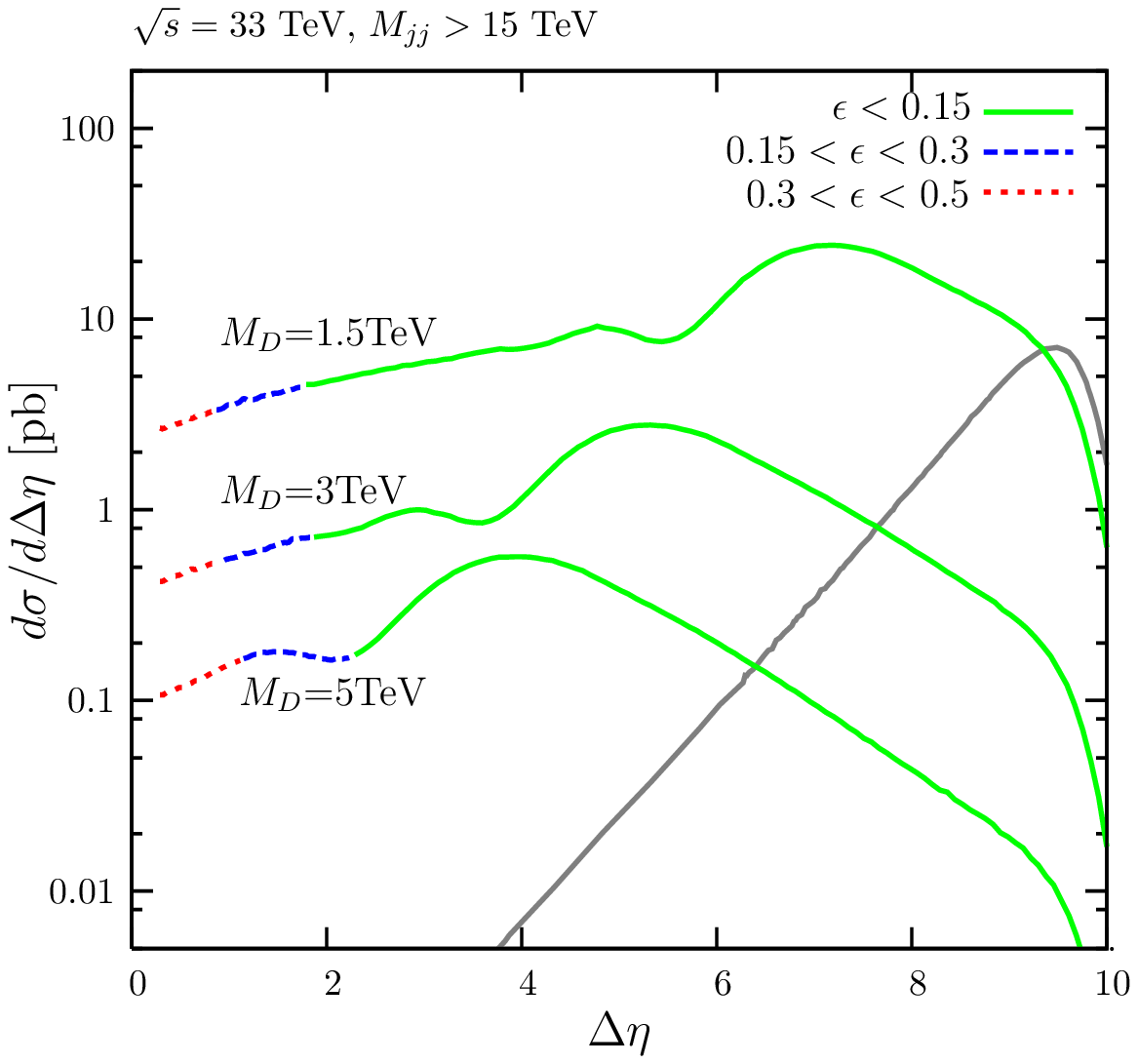}
\caption{Differential cross section $d\sigma/d\Delta\eta$ for $n=6$, at $\sqrt{s}=$33~TeV and $M_{JJ}^{min}=$15~TeV.}
\label{sb33}
\end{minipage}
\end{figure}

On increasing the centre-of-mass energy, the calculability region gets wider as shown by the value of the parameter $\epsilon$ as a function of kinematics. Here, we calculate the differential cross section $d\sigma/d\Delta\eta$ as a function of $\Delta\eta$ using the parameter $\epsilon$ to visualise where the amplitude is reliably calculable. The $p_T$ and $\eta$ cuts of Section 2 are imposed here to account for detector acceptance. We note that in this section we restrict ourselves to calculating the signal and background for positive values of $\Delta\eta$, rejecting the region of $\Delta\eta<$0 which corresponds to large momentum transfer scattering which cannot be calculated within the eikonal approximation. The contribution of this region is expected to be sufficiently small. The value of $\epsilon$ is calculated at each value of $\Delta\eta$ using the invariant dijet mass cut squared in place of $s$ in Eq.~\ref{epsiloneq}, calculated for $M_D$=1.5, 3, 5~TeV with $M_{JJ}^{min}=9$~TeV at the LHC in Fig.~\ref{sb14}. The signal lines have three colours depending on the value of $\epsilon$: green (solid) for $\epsilon <$0.15, blue (dashed) for $0.15<\epsilon<0.3$ and red (dotted) for $0.3<\epsilon<0.5$. The plots do not extend into regions where $\epsilon>0.5$ as there the calculation is not reliable.
Calculability improves (i.e., $\epsilon$ becomes smaller) by increasing the centre-of-mass energy as we can then probe deeply into the transplanckian region. This is illustrated in Fig.~\ref{sb33}, where the dijet invariant mass cut has been raised to 15~TeV, an energy not accessible at present LHC energies. Now the $M_D$=5~TeV line has turned green ($\epsilon<0.15$) which shows that higher $M_D$ scales can now be probed reliably, with the signal to background ratio being larger and the green (solid) region extended for all $M_D$. 

\begin{figure}
\begin{minipage}[b]{0.5\linewidth} 
\includegraphics[trim=1cm 0 0 0,width=7.35cm,height=7cm]{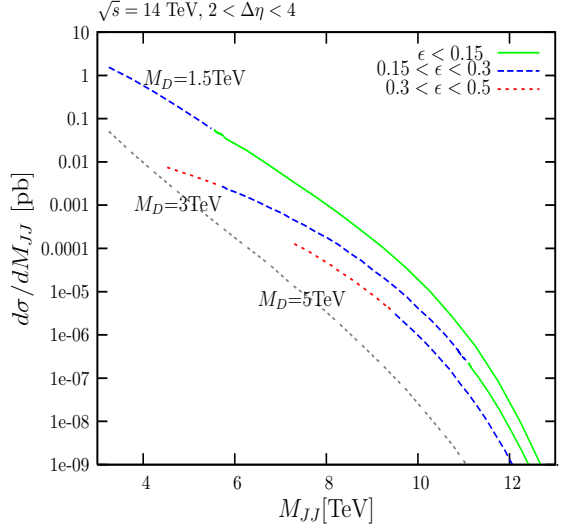}
\end{minipage}
\hspace{0.2cm} 
\begin{minipage}[b]{0.5\linewidth}
\includegraphics[trim=1cm 0 0 0,width=7.35cm,height=7cm]{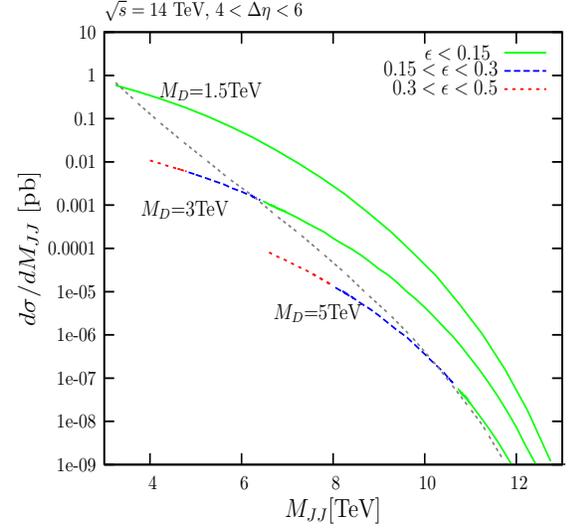}
\end{minipage}
\caption{Differential cross section $d\sigma/dM_{JJ}$ for $n=6$, as a function of $M_{JJ}$ for $2<\Delta\eta<4$ and $4<\Delta\eta<6$ at $\sqrt{s}$=14~TeV. }
\label{mjj14}
\end{figure}

\begin{figure}
\begin{minipage}[b]{0.5\linewidth} 
\includegraphics[trim=1cm 0 0 0,width=7.4cm,height=7cm]{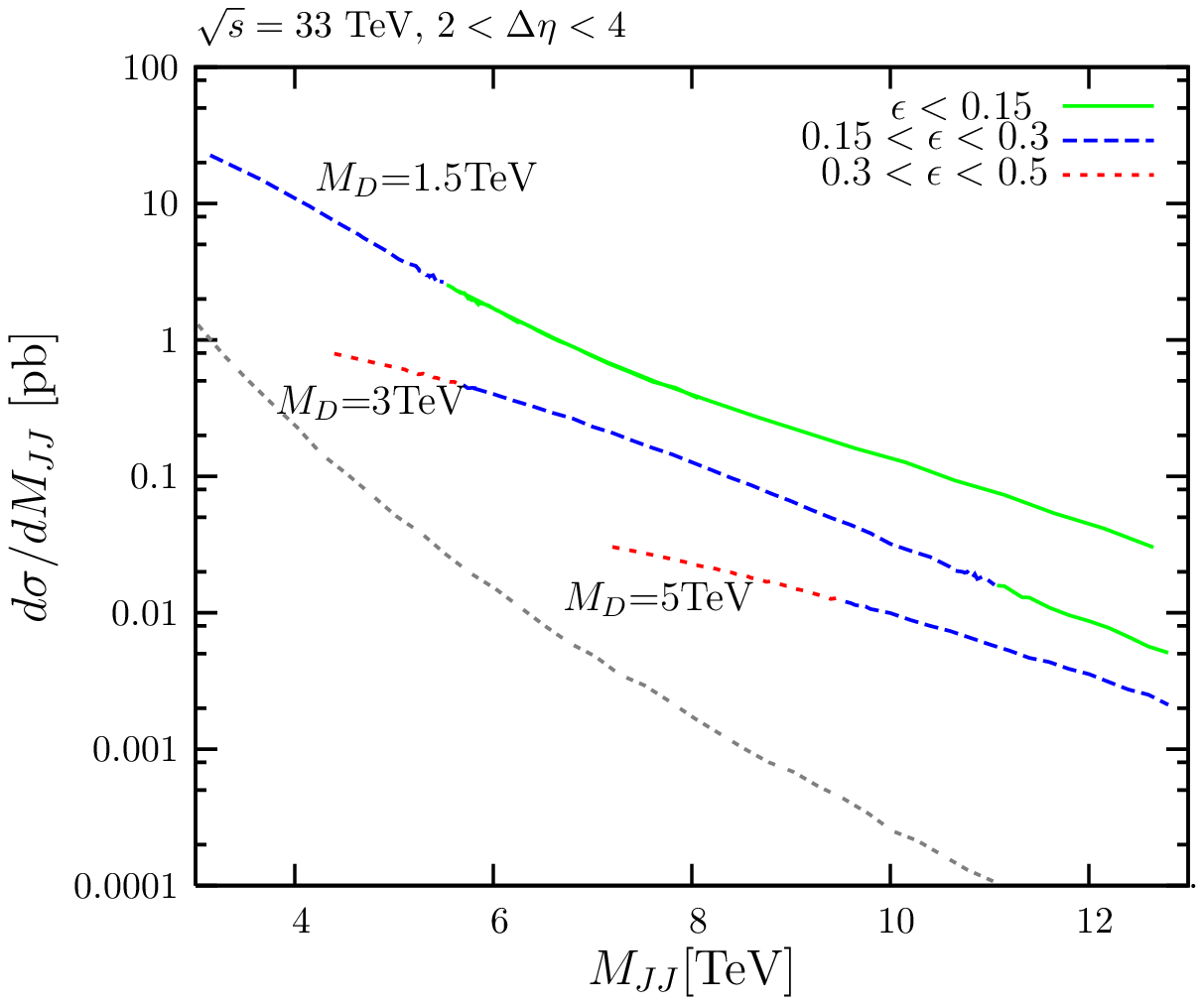}
\end{minipage}
\hspace{0.2cm} 
\begin{minipage}[b]{0.5\linewidth}
\includegraphics[trim=1cm 0 0 0,width=7.4cm,height=7cm]{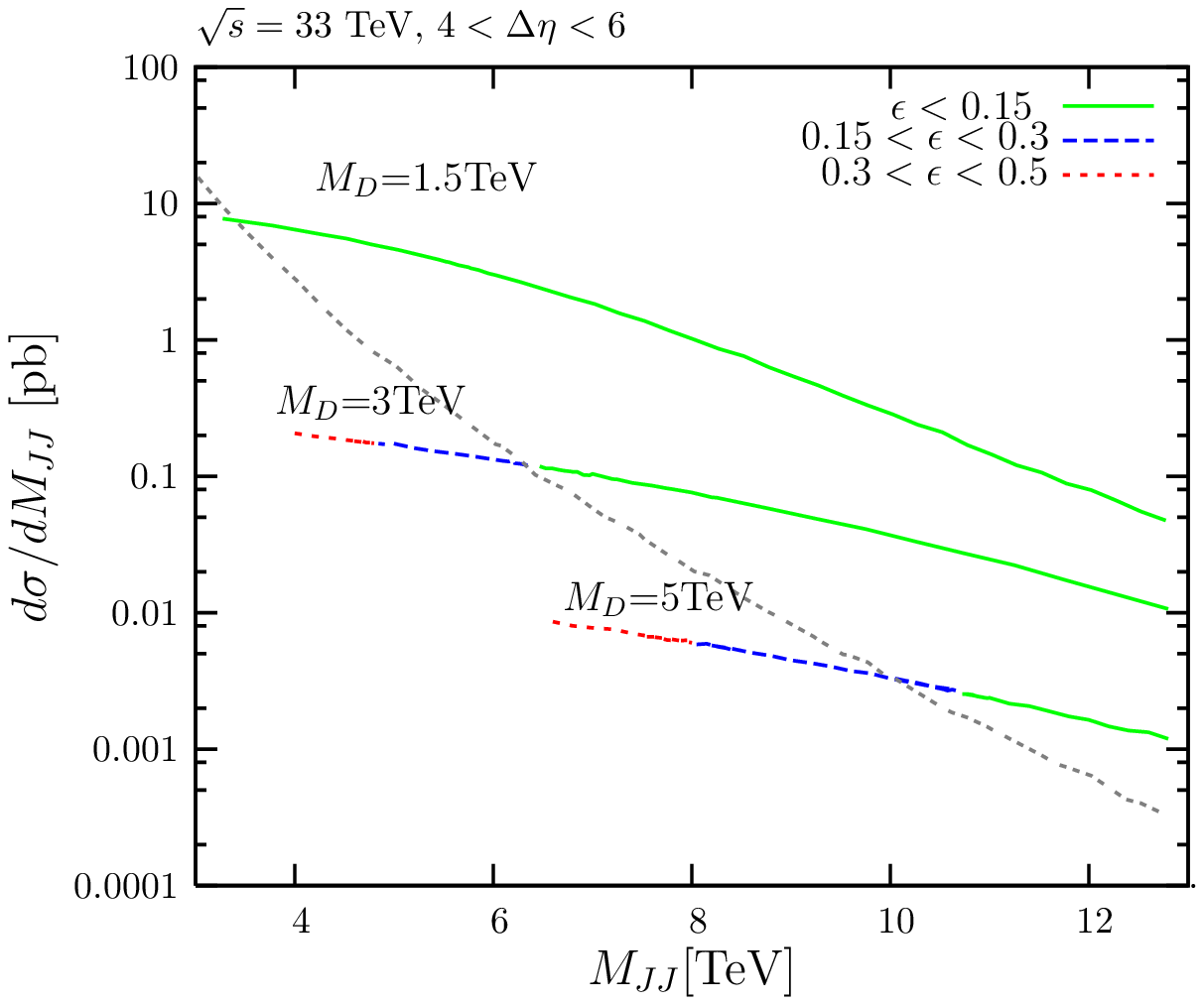}
\end{minipage}
\caption{Differential cross section $d\sigma/dM_{JJ}$ for $n=6$, as a function of $M_{JJ}$ for $2<\Delta\eta<4$ and $4<\Delta\eta<6$ at $\sqrt{s}$=33~TeV.}
\label{mjj33}
\end{figure}

In a similar way, computability can be visualised for the differential cross section $d\sigma/dM_{JJ}$. Keeping the same $p_T$ and individual jet $\eta$ cuts of Section 3, we calculate $d\sigma/dM_{JJ}$ for two different $\Delta\eta$ bins: $2<\Delta\eta<4$ and $4<\Delta\eta<6$. Here $\epsilon$ is calculated using the minimum value of $\Delta\eta$ in each bin to give the most  stringent limit on computability. The cross section is calculated for $M_D$=1.5, 3, 5~TeV at both 14 and 33~TeV for the same range of $M_{JJ}$ values and shown in Fig.~\ref{mjj14} and Fig.~\ref{mjj33} respectively. We note that using the minimum value of $\Delta\eta$ in each bin  to calculate $\epsilon$ for $d\sigma/dM_{JJ}$ gives generally stronger limits for computability compared to using the $M_{JJ}$ cut for $d\sigma/d\Delta\eta$, as $d\sigma/dM_{JJ}$ is a rapidly decreasing function of $M_{JJ}$ whereas $d\sigma/d\Delta\eta$ is not a decreasing function of $\Delta\eta$. In all plots we show the QCD background, as we need to keep in mind that in addition to computability, the signal to background ratio is crucial for any experimental search. In the plots, we see again that setting a higher $\Delta\eta$ cut improves the reliability of the calculation but decreases the signal to background ratio. Thus, an intermediate region of $\Delta\eta$ is ideal. Again increasing the centre-of-mass energy gives better signal to background ratio and a wider calculability region.

\section{Conclusions}
We have studied transplanckian scattering at the LHC within the eikonal approximation. In the kinematical limits of high rapidity separation between the two jets and high centre-of-mass energy, we expect the distinctive features of the signal to allow us to distinguish it from the QCD background, which we have also calculated and presented. At the 14~TeV LHC, given the current limits on the fundamental mass scale, it is only marginally possible to reach the transplanckian regime. We have studied how the signal and background are affected by the fundamental mass scale, the scale choice for the PDFs and the approach needed to maximise the signal to background ratio. We also considered the computability of the eikonal signal and considered the corrections to the eikonal amplitude to establish the region of parameter space where the eikonal amplitude is most reliably calculable. 

We considered how a possible LHC energy upgrade to 33~TeV centre-of-mass energy will improve the prospect of observing transplanckian scattering at the LHC. Our studies have shown that even with $M_D\geq 5$ TeV there is in principle a calculable transplanckian kinematic region with good signal rate ($\sim$ fb) and favourable signal to background ratio.  This constitutes another illustration of how increasing the energy is typically more important in most ``gravity phenomenology" than increasing luminosity.  Thus, from this perspective the HE LHC presents qualitative improvement in covering the parameter space of extra dimensional theories with its prospects for probing the cisplanckian, planckian and transplanckian regimes, whose signals could be determined through control of kinematic variables.

\acknowledgments{E.V. acknowledges financial support from the UK Science and Technology Facilities Council.}

\end{document}